\documentclass[acp, manuscript]{copernicus}

\usepackage{epstopdf,ulem}
\usepackage{mathrsfs} %for relative humidity H
\usepackage{booktabs}
\usepackage{varwidth}
\usepackage{xcolor}

\usepackage{setspace}

\usepackage{amsmath,amsfonts,amssymb}
\usepackage{bm,latexsym,euscript,textcomp}
\usepackage{graphicx,color,graphics,epsf,dcolumn}
\usepackage{natbib}

\begin{document}
\nolinenumbers

\title{Comments on ``An evaluation of hurricane superintensity in
axisymmetric numerical models" by Rapha\"el Rousseau-Rizzi
and Kerry Emanuel }

% \Author[affil]{given_name}{surname}

\Author[1,2]{Anastassia M.}{Makarieva}
\Author[1]{Andrei V.}{Nefiodov}
\Author[3]{Douglas}{Sheil}
\Author[4]{Antonio Donato}{Nobre}
\Author[5]{Alexander V. }{Chikunov}
\Author[6]{G\"{u}nter}{Plunien}
\Author[2]{Bai-Lian}{Li}

\affil[1]{Theoretical Physics Division, Petersburg Nuclear Physics Institute, Gatchina  188300, St.~Petersburg, Russia}
\affil[2]{USDA-China MOST Joint Research Center for AgroEcology and Sustainability, University of California, Riverside 92521-0124, USA}
\affil[3]{Faculty of Environmental Sciences and Natural Resource Management, Norwegian University of Life Sciences, \AA s, Norway}
\affil[4]{Centro de Ci\^{e}ncia do Sistema Terrestre INPE, S\~{a}o Jos\'{e} dos Campos, S\~{a}o Paulo  12227-010, Brazil}
\affil[5]{Princeton Institute of Life Sciences, Princeton, New Jersey 08540, USA}
\affil[6]{Institut f\"{u}r Theoretische Physik, Technische Universit\"{a}t Dresden,  D-01062 Dresden, Germany}

%% The [] brackets identify the author with the corresponding affiliation. 1, 2, 3, etc. should be inserted.

\runningtitle{Comments on ``An evaluation of hurricane superintensity in
axisymmetric numerical models"}

\runningauthor{Makarieva et al.}

\correspondence{A. M. Makarieva (ammakarieva@gmail.com)}

\received{}
\pubdiscuss{} %% only important for two-stage journals
\revised{}
\accepted{}
\published{}

%% These dates will be inserted by Copernicus Publications during the typesetting process.

\firstpage{1}

\maketitle

\begin{abstract}
In a recent paper \citet{em19} presented a derivation of an upper limit on maximum hurricane velocity at the surface.
This derivation was based on a consideration of an infinitely narrow (differential) Carnot cycle with the warmer isotherm
at the point of the maximum wind velocity. Here we show that this derivation neglected a significant term describing the kinetic energy change in the outflow. Additionally, we highlight the importance of a proper accounting for the power needed to lift liquid water. Finally, we provide a revision to the formula for surface fluxes of heat and momentum showing that, if we accept the assumptions adopted by \citet{em19}, the resulting velocity estimate does not depend on the flux of sensible heat.
\end{abstract}

%\Large
\introduction  %% \introduction[modified heading if necessary]

\label{intro}

\citet{em19} (hereafter RE) presented a new derivation of surface potential intensity (PI). They based this on consideration
of an infinitely narrow Carnot cycle in the vicinity of maximum wind speed.  Cyclonic storms, especially powerful hurricanes, represent a serious threat to human lives in many regions of the world but our ability to understand and anticipate these storms remains incomplete.  Finding theoretical constraints on maximum hurricane velocities is an important goal, and we welcome the new work and its subsequent discussion  by \citet{ms20} and \citet{em20}.

In their derivation, RE considered a configuration of closed air streamlines shown in Fig.~1a. \citet{ms20}  pointed out that such a configuration is not realistic and that the actual streamline contours should be enclosed within one another as shown in Fig.~1b.  \citet{em20} replied that, for their derivation to be valid,  the air does not have to actually move as shown in Fig.~1a; in particular, it does not have to actually move along the warmer isotherm from $\rm B^\prime$ to $\rm B$.
Here we reconsider these arguments and show that the streamline configuration does matter for the derivation of RE and that accounting for realistic air motion \citep[see Fig.~1 of][]{makarieva18b} leads to the appearance of a significant
term characterizing the outflow region.

Furthermore, \citet{em20} clarified that RE's derivation of surface PI assumed reversible thermodynamics whereby the total water content of air parcel ($q_t$) is not supposed to change. RE neglected the last term in their Eq.~(13). This term is proportional to $dq_t/dt$ and, in the general case, describes the power needed to lift liquid water.
If $q_t$ is constant, this term should be zero and could be discarded as RE did. However, 
\citet{sabuwala15} estimated this term in the real atmosphere to be significant leading 
to up to a 30\% reduction of PI. This would make RE's assumption of reversible thermodynamics of limited
relevance for real PIs. We discuss the other available estimate of the power needed to lift liquid water by \citet{makarieva18},
which revises the estimate of \citet{sabuwala15} and  shows the corresponding term to be small thus restoring the practical relevance of RE's derivation in this aspect.

Finally, \citet{ms20} question how RE's resulting formula for surface PI, RE's Eq.~(15), is obtained from the consideration of the differential Carnot cycle summarized in RE's Eq.~(14).  Indeed the expressions in those equations have different units, W~m$^{-3}$ for the integrand in RE's Eq.~(14) versus W~m$^{-2}$ in RE's Eq.~(15). In their reply to \citet{ms20},     \citet{em20} did not provide  an explicit derivation to show how RE's Eq.~(15) can be derived from RE's Eq.~(14).
We clarify the physical assumption behind this transition and show that, once the definition of entropy is explicitly considered for the warmer isotherm of the differential Carnot cycle, the resulting revised formula relates surface PI to latent heat flux only and PI is independent of the flux of sensible heat.

\section{Integrals over closed contours}
\label{oint} 

RE noted that the material derivative of pressure, $dp/dt$, enters both
the definition of the material derivative of moist entropy $ds/dt$ (Eq.~(9) of RE) and 
the scalar product of the equation of motion with the three-dimensional velocity $\mathbf{V}$ (Eq.~(10) of RE):
\begin{gather}
\label{eq9}
T\frac{ds}{dt}=(c_{pd} + c_l q_t)\frac{dT}{dt} + \frac{d(L_v q)}{dt} - (1 + q_t)\alpha \frac{dp}{dt} - R_vT \ln (\mathscr{H}) \frac{dq_t}{dt}, \\
\label{eq10}
\frac12\frac{d |\mathbf{V}|^2}{dt}= -\alpha \frac{dp}{dt} + \mathbf{F}\cdot\mathbf{V} - wg.
\end{gather}
The standard notations follow RE.

RE eliminated $dp/dt$ between the two equations and applied the resulting relationship for $Tds/dt$ (their Eq.~11) 
to two closed air trajectories of hurricane air, $\rm ABCDA$ and $\rm AB^{\prime}C^{\prime}DA$ (Fig. 1a). Subtracting the two contour integrals of $Tds/dt$ from one another, RE concluded that, along the infinitesimal inner loop
$\rm B^{\prime}BCC^{\prime}B^{\prime}$, the contour integrals of $Tds/dt$  and of the friction power $\mathbf{F}\cdot\mathbf{V}$ ($\mathbf{F}$ is the ``frictional source of momentum")  coincide, as summarized by RE's Eq.~(14).
To arrive at their Eq.~(14), RE assumed that the integrals over closed contours of any $dX/dt$ (in particular, of $X = |\mathbf{V}|^2$) are zero.

\begin{figure*}[tbp]
%\vspace{-0.5 cm}
\begin{minipage}[p]{1\textwidth}
\centering\includegraphics[width=0.8\textwidth,angle=0,clip]{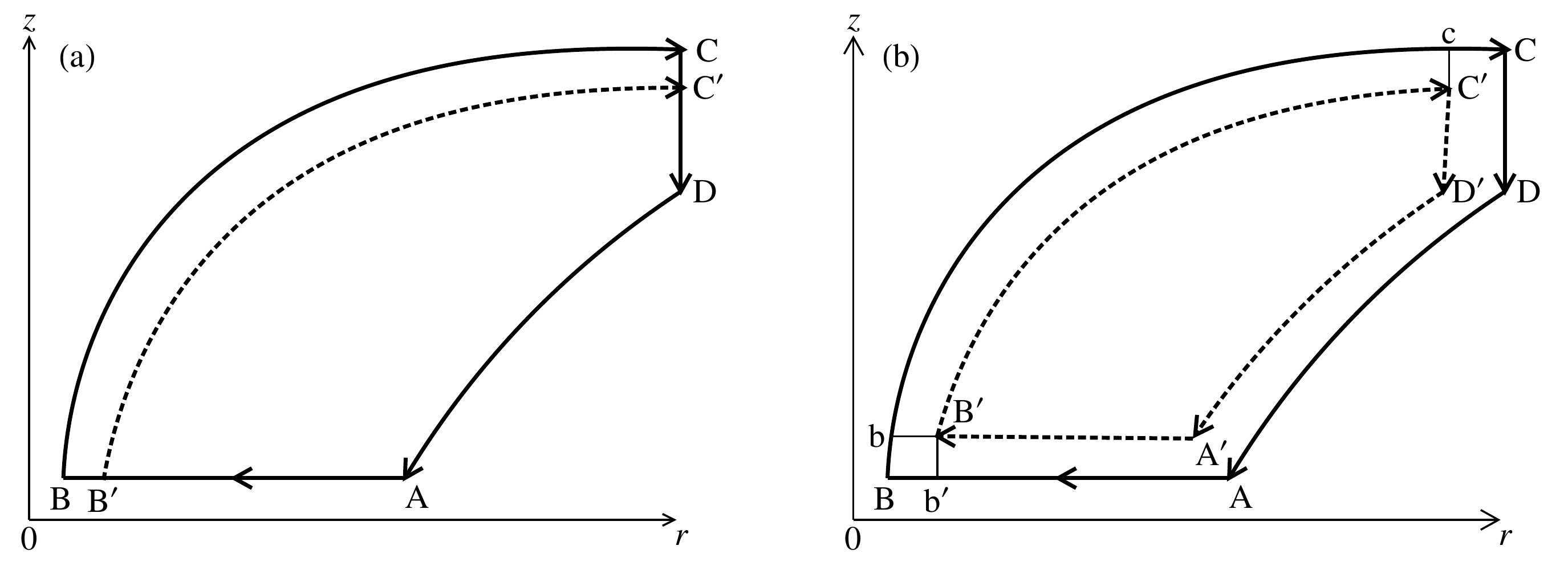}
\end{minipage}
\caption{%\Large
(a) Configuration of air streamlines considerd by RE (their Fig.~1).
(b) A realistic configuration of air streamlines. The axes of $z$ and $r$ correspond to the vertical and radial variables, respectively. Points $\rm B$ and $\rm B^{\prime}$ are infinitely close and chosen at the point of maximum wind.
}
\label{fig1}
\end{figure*}

\citet{ms20} pointed out that RE did not indicate the integration variables in their integrals. In their reply \citet{em20} did not clarify this omission. In a steady state, material derivative  is defined as follows:
\begin{equation}
\label{md}
\frac{dX}{dt} \equiv \mathbf{V} \cdot \nabla X.
\end{equation}
The integral of the material derivative over a closed contour is equal to zero only if the contour is a streamline.
Moving with the parcel over the contour and taking into account that $\mathbf{V} = d\mathbf{l}/dt$,
where $d\mathbf{l}$ is directed along the streamline, one has for any scalar quantity $X$:	
\begin{equation}
\label{ointe}
\oint \frac{dX}{dt} dt = \oint \mathbf{V} \cdot \nabla X dt = \oint \frac{d\mathbf{l}}{dt} \cdot \nabla X dt = \oint d\mathbf{l} \cdot \nabla X = \oint dX = 0 .
\end{equation}
If the closed contour is not  a streamline, the integral is not zero.

Since streamlines are parallel to velocity and since velocity is unambiguously defined at each point,  there cannot be two different streamlines emanating from or entering a single point, like B$^{\prime}$ and C$^{\prime}$ in Fig. 1 of RE (Fig. 1a).
The configuration shown in Fig.~1a is therefore impossible: the inner ``differential" loop  $\rm B^\prime BCC ^\prime B^\prime$ is not composed of streamlines or their parts. A realistic configuration of streamlines is shown in Fig. 1b.
A consistent application of the thermodynamic and dynamic equations (\ref{eq9}) and (\ref{eq10}) to an infinitely
narrow  loop $\rm B^{\prime}bcC^{\prime}B^{\prime}$ in the vicinity of maximum wind requires an explicit consideration
of two infinitely close streamlines, $\rm B^{\prime}C^{\prime}$ and $\rm b^{\prime}Bc$ (Fig. 1b).

The point about the unrealistic configuration of streamlines was put forward by \citet{ms20}. However, they did not specify
the consequences of this assumption for RE's derivations. As we discuss below, the main implication is that the integral of $d|\mathbf{V}|^2/dt$ over $dt$ for the inner ``differential" cycle is not zero.

There is an essential difference between Eqs. (\ref{eq9}) and (\ref{eq10}) (Eqs. (9) and (10) of RE).
Equation~(\ref{eq9}) represents a definition of entropy, to which the operator $d/dt$ has been applied.
Instead of $d/dt$, one can apply to the definition of entropy an operator $d\mathbf {l} \cdot \nabla$,
where $d\mathbf{l}$ is an arbitrary vector (not necessarily parallel to velocity vector).
Formally it is equivalent to replacing $d/dt$ in Eq.~(\ref{eq9})  with $d\mathbf {l} \cdot \nabla$. 

If one assumes, following RE, that the inner loop $\rm B^{\prime}bcC^{\prime}B^{\prime}$ is a Carnot cycle, with two adiabats $\rm bc$ and $\rm B^{\prime}C^{\prime}$ and two isotherms $\rm B^{\prime}b$ and $\rm cC^{\prime}$,
one can write an analogue of Eq.~(\ref{eq9}) for this loop as
\begin{equation}\label{wc}
-\oint (1 + q_t)\alpha dp = \oint T ds^* = \varepsilon_C \int_{\rm B^\prime}^{\rm b} \!\!\delta Q .
\end{equation}
Here $q_t$ is the mixing ratio for total water,  $\alpha=1/\rho$, $\rho$ is air density, $s^*$ is saturated moist entropy corresponding to relative humidity  $\mathscr{H} = 1$, $\delta Q = Tds^*$ is heat increment, and $\varepsilon_C = 
(T_{\rm b}- T_{\rm c})/ T_{\rm b}$ is Carnot efficiency. The temperatures $T_{\rm b}$ and $T_{\rm c}$ correspond to the points b and c, respectively (see Fig.~1b). The first integral in Eq.~(\ref{wc}) represents work of the cycle per unit mass of dry air.

\citet{ms20} criticized RE for not accounting for ice melting.  However, Eq.~(\ref{wc}) is valid for any Carnot cycle, whether or not it includes ice. The Carnot cycle efficiency does not depend on the latent heat of vaporization (sublimation, melting)
or heat capacity $c_l$ of liquid water that all enter the definition of entropy, see Eq.~(\ref{eq9}).  So if the cycle is reversible, the result is uninfluenced by ice.

Importantly, Eq.~(\ref{wc})  relates state variables and is thus valid independent of whether the air {\it actually}
moves along the considered closed contour or not. Indeed,  \citet{em20} pointed out that, for the thermodynamics, it is not essential whether the air actually moves from point $\rm B^\prime$ to $\rm B$.

But Eq.~(\ref{eq10}), unlike Eq.~(\ref{eq9}), is not a definition; it is a law of motion. Replacing $d/dt$ with $d\mathbf {l} \cdot \nabla$ and $\mathbf{V}$ with $d\mathbf{l}$ in Eq.~({\ref{eq10})  is only possible if $d\mathbf{l}$  is part of a streamline. This replacement transforms Eq.~(\ref{eq10}) into the Bernoulli equation that is only valid on a streamline:
\begin{equation}
\label{B}
\frac12 d |\mathbf{V}|^2 = -\alpha dp + \mathbf{F}\cdot d\mathbf{l} - gdz.
\end{equation}
Assuming, again following RE, that along the streamlines $\rm B^\prime C^\prime$ and $\rm bc$ there
is no friction, $\mathbf{F} = 0$, and applying Eq.~(\ref{B}) to these two streamlines,
one can express the integral of $\alpha dp$ over the closed contour $\rm B^{\prime}bcC^{\prime}B^{\prime}$  as
\begin{gather}\label{cc}
-\oint \alpha dp = -\int_{\rm B^\prime}^{\rm b} \!\!\alpha dp  -\int_{\rm c}^{\rm C^\prime} \!\!\!\alpha dp
+ \frac{V_{\rm c}^2- V_{\rm C^\prime}^2}{2} - \frac{V_{\rm b}^2- V_{\rm B^\prime}^2}{2} + g(z_{\rm c} - z_{\rm C^\prime})
=-\int_{\rm B^\prime}^{\rm b} \!\!\alpha dp 
- \frac{V_{\rm b}^2- V_{\rm B^\prime}^2}{2} + \frac{V_{\rm c} ^2- V_{\rm C^\prime}^2}{2}.
\end{gather}
In the last equality of Eq.~(\ref{cc}) the hydrostatic equilibrium $\alpha \partial p/\partial z = -g$ was applied along the vertical path $\rm c C^\prime$.

Eliminating $\displaystyle\oint \alpha dp$ between Eqs.~(\ref{wc})  and (\ref{cc}) -- this procedure is analogous to RE
eliminating $dp/dt$ between their Eqs.~(9) and (10) -- yields:
\begin{equation}\label{prom}
\oint T ds^*  = 
- \int_{\rm B^\prime}^{\rm b} \!\!\alpha dp 
- \frac{V_{\rm b}^2- V_{\rm B^\prime}^2}{2} + \frac{V_{\rm c}^2- V_{\rm C^\prime}^2}{2} - \oint q_t \alpha dp.
\end{equation}

Assuming hydrostatic equilibrium at point $\rm B$, assuming that we are in the region of maximum wind velocity 
($d |\mathbf{V}|^2=0$) and considering the streamline $\rm b^\prime B b$ in the limit $\rm b^\prime \to B^\prime$ of two infinitely close streamlines, with use of the  Bernoulli equation (\ref{B}) one obtains: 
\begin{equation}\label{lim}
\int_{\rm b^\prime}^{\rm b} \!\!\left(\alpha \frac{\partial p}{\partial r} dr +  \alpha \frac{\partial p}{\partial z} dz+ g dz\right)
=  \int_{\rm b^\prime}^{\rm b} \!\!\mathbf{F}\cdot d\mathbf{l} = \int_{\rm b^\prime}^{\rm B} \!\!\alpha dp  \underset{\rm b^\prime \to B^\prime}{\longrightarrow} \int_{\rm B^\prime}^{\rm b} \!\!\alpha dp .
\end{equation}
Using Eq.~(\ref{lim}) one can write Eq.~(\ref{prom}) as
\begin{equation}\label{fin}
\lim_{\rm b^\prime \to B^\prime}\oint T ds^*  =
-\int_{\rm b^\prime}^{\rm b} \!\!\mathbf{F}\cdot d\mathbf{l} 
- \oint q_t \alpha dp + \frac{V_{\rm c}^2- V_{\rm C^\prime}^2}{2}.
\end{equation}
The first integral in the right-hand side is taken along the streamline $\rm b^\prime B b$. This equation is analogous to RE's Eq.~(13) but it correctly takes into account  the change of kinetic energy in the outflow region. \citet[][see their Appendix~C]{makarieva18b} showed that the last term  in Eq.~(\ref{fin}), neglected by RE, is significant when the outflow radius $r_{\rm C}$ (defined in the above derivation as the radius where the ascending air reaches the tropopause, i.e. the vertical isotherm $\partial T/\partial z = 0$) is close to the radius of maximum wind, $r_{\rm C} \sim r_{\rm B}$. The implication is that RE's derivation should be less relevant when the rising air reaches the tropopause  close to the radius of maximum winds.

\section{The power to lift water}

The second term in the right-hand side of Eq.~(\ref{fin}) represents work associated  with lifting liquid water and changing its kinetic energy.  For a closed streamline with $\mathbf{F} = 0$ one has from the Bernoulli equation 
\begin{equation}\label{qt}
-\oint q_t \alpha dp = - \oint dq_t \left(\frac{1}{2} |\mathbf{V}|^2 + gz\right)
\end{equation}
\citep[for details see][]{makarieva18b}.
This term corresponds to the second term in the right-hand side of Eq.~(13) of RE who
wrote: ``The last term in Eq.~(13) represents the irreversible entropy loss associated with lifting water
mass against gravity and changing its kinetic energy. It is quantitatively small compared to the other terms
in Eq.~(13) and we henceforth neglect it."

\citet{em20} clarified that in their derivations RE assumed  reversible thermodynamics under which $q_t$ should not change. In this case $dq_t = 0$ and the corresponding term is zero. In the real atmosphere condensed moisture is removed from the rising air parcels by precipitation and $q_t$ is not constant. If the corresponding term in RE's Eq.~(13) were large,
neglecting it would limit the application of their results to the real atmosphere.

RE did not support their statement about the term being quantitatively small.  \citet{em18} quoted two studies that evaluated how the estimate of the potential intensity of tropical cyclones can be changed by accounting for lifting water, those of \citet{sabuwala15} and \citet{makarieva18}. Among the two, \citet{sabuwala15} indicated that lifting water can {\it decrease} potential intensity by as much as thirty per cent. This runs counter to the statement of RE that the corresponding term is ``quantitatively small"; RE did not quote \citet{sabuwala15}.

\citet{makarieva18}, on the other hand, showed that the analysis of \citet{sabuwala15}
was in error and reported the first ever, to our knowledge, estimate of 
the gravitational power of precipitation (lifting water) in tropical cyclones.
In making use of the size of this term, an essential aspect of their analyses, RE did not cite where this result had been established.

\section{Relationship between surface and volume fluxes}

\citet{ms20} noted that RE did not explain how their final Eq.~(15) relating  surface fluxes of heat and turbulent friction was obtained from their Eq.~(14). Here we briefly clarify and revise such a derivation. Following RE, we neglect the last two terms in Eq.~(\ref{fin}). Then, by using the last equality in Eq.~(\ref{wc}), Eq.~(\ref{fin}) can be written as
\begin{equation}\label{finn}
\varepsilon_C \lim_{\rm b^\prime \to B^\prime} \int_{\rm B^\prime}^{\rm b}\!\! \delta Q = - \int_{\rm b^\prime}^{\rm b} \!\!\mathbf{F}\cdot d\mathbf{l}.
\end{equation}
Under the additional assumption that the adiabats below $\rm B^\prime b$ are vertical, such
that only horizontal air motion is associated with heat input into the air parcel, the above
equation can be interpreted as follows: any time an air parcel moves from $\rm b^\prime$ to $\rm b$,
the work of the friction force $-\mathbf{F}\cdot d\mathbf{l}$ equals $\varepsilon_C$ times the heat the air parcel receives.
The units of these variables are joule per kilogram of dry air.

{\it Assuming} that the same ratio characterizes the surface fluxes of heat and momentum,
one can conclude that the surface flux of momentum $D=\rho C_{D} |\mathbf{V}|^3$
equals $\varepsilon_C$ times the surface flux of heat $J= \rho C_{k} |\mathbf{V}| (k^*_s - k)$ (dimension W~m$^{-2}$):
\begin{equation}\label{rat}
\varepsilon_C = \frac{ -\mathbf{F}\cdot d\mathbf{l} }{\delta Q} =  \frac{D}{J} , 
\end{equation}
where $C_{D}$ and $C_{k}$ are surface exchange coefficients, $k^*_s$ is saturated enthalpy at surface temperature, and $k$ is the actual enthalpy in the near-surface layer. This assumption is key for RE's derivation:   RE's Eq.~(15) could not have been obtained otherwise from their Eq.~(14).  A similar  relationship between surface-specific and volume-specific energy fluxes was adopted by \citet[][their Eq.~(18)]{makarieva18b}. We consider it reasonable.
Indeed, if air motion above the boundary layer is adiabatic, all the surface heat flux should be accommodated into the volume of air parcels moving above the surface within the boundary layer.

In their Eq.~(15), RE formulated the surface heat flux as the flux determined by the difference in enthalpies plus the so-called dissipative heating equal to the flux of momentum $D$. However, inspection of Eq.~(\ref{finn}) reveals that the heat increment $\delta Q$ already contains the work of the friction force, which cannot be accommodated into the heat input yet another time, see Eq.~(\ref{lim}):
\begin{equation}
\label{dQ}
\int_{\rm B^\prime}^{\rm b} \!\!\delta Q = \int_{\rm B^\prime}^{\rm b} \!\!(L_v dq - \alpha dp) 
\underset{\rm b^\prime \to B^\prime}{\longrightarrow}
\int_{\rm B^\prime}^{\rm b} \!\!L_v \frac{\partial q}{\partial r}dr - \int_{\rm b^\prime}^{\rm b} \!\!\mathbf{F}\cdot d\mathbf{l}.
\end{equation}
Here $L_v$ is the latent heat of vaporization and it is assumed that $q_t \ll 1$. Heat input from the surface arrives to within an air parcel in two forms: latent heat $L_v dq$ that is associated with change $dq$ of the water-vapor mixing ratio  and sensible heat $-\alpha dp$ that the air parcel accommodates by expansion (to remain isothermal). Thus, if, as Eq.~(\ref{lim}) prescribes, $-\alpha dp$ is equal to the work of friction $-\mathbf{F}\cdot d\mathbf{l}$,  and if all this work of friction dissipates to heat within the air parcel, no more heat can be accommodated in sensible form from the ocean.  As we previously argued, there is no enhancement of hurricane intensity due to dissipative heating \citep[see discussions by][]{dhe10,bister11,bejan19,bejan20}.

Using Eq.~(\ref{dQ})  the assumed relationship Eq.~(\ref{rat}) can be re-written as
\begin{equation}\label{Lrat}
\frac{\varepsilon_C}{1 - \varepsilon_C} = \frac{-\mathbf{F}\cdot d\mathbf{l}}{L_v (\partial q/\partial r) dr}=\frac{-\mathbf{F}\cdot \mathbf{V}}{L_v (\partial q/\partial r) u}=\frac{D}{J_L},
\end{equation}
where $J_L= \rho C_{k} |\mathbf{V}| L_v ( q^*_s - q)$ is the surface flux of latent heat,   $q^*_s$ is the saturated vapor mixing ratio at sea surface temperature,  $q$  is the vapor mixing ratio in the near-surface layer, and $u \equiv dr/dt$ is radial velocity.  The third ratio in Eq.~(\ref{Lrat}) is the ratio of the volume-specific fluxes of frictional dissipation and  heat consumption  (dimension W~m$^{-3}$).

One essential point is that  RE assumed $\mathbf{F} = 0$ everywhere above $\rm B^\prime B$ on their contour (Fig.~1a).  This assumption can be valid if point $\rm B^\prime$ is chosen sufficiently close to the top of the boundary layer, above which  friction is commonly assumed to be negligible.  If point $\rm B^\prime$ is located at the top of the boundary layer, such that friction tends to zero, $|\mathbf{F}| \to 0$,   in the considered limit $\rm b^\prime \to B^\prime$, 
the second and third ratios in Eq.~(\ref{Lrat}) can nevertheless remain finite (not zero)  if the air leaving the boundary layer has zero radial velocity $u$, i.e. if $|d\mathbf{l}/dr |\to \infty$ as $\rm b^\prime \to B^\prime$.

In the view of Eq.~(\ref{Lrat}) the revised version of RE's Eq.~(16) becomes
\begin{equation}
|\mathbf{V}|^2=   \frac{T_{\rm b} - T_{\rm c}}{T_{\rm c}}   \frac{C_{k}}{C_{D}} L_v(q^*_s - q ) ,
\end{equation}
where $T_{\rm b}$ and $T_{\rm c}$ are temperatures of the warm and cold isotherms, respectively. 
The wind speed $|\mathbf{V}|$ at the radius of maximum winds,   vapor  mixing ratio $q$, and the exchange coefficients $C_{k}$ and $C_{D}$  pertain to 10-m altitude. This expression relates latent heat input to losses due to friction and avoids consideration of the temperature difference between the sea surface and the adjacent air,
as explained in detail by \citet{makarieva18b}.

\bibliographystyle{copernicus}
%\bibliography{met-refs}

\end{document}